\documentclass[onecolumn,12pt]{revtex4}

\usepackage{amsmath}
\usepackage{amsbsy}

\usepackage{amsmath,amssymb,subfigure}
\usepackage{graphicx}
\usepackage{epsfig}
\usepackage{color}

\newcommand{\beq}{\begin{equation}}
\newcommand{\eeq}{\end{equation}}
\newcommand{\beqa}{\begin{eqnarray}}
\newcommand{\eeqa}{\end{eqnarray}}
\newcommand{\om}{\Omega_m}

\newcommand{\ls}{\mathrel{\raise0.27ex\hbox{$<$}\kern-0.70em \lower0.71ex\hbox{{
$\scriptstyle \sim$}}}}

\begin{document} 

\title{Einstein's Other Gravity and the Acceleration of the Universe} 
\author{Eric V.\ Linder} 
\affiliation{Berkeley Lab \& University of California, Berkeley, CA 94720 
USA\\ 
Institute for the Early Universe, Ewha Womans University, Seoul 120-750 Korea 
}

\date{\today}

%%%%%%%%%%%%%%%%%%%%%%%%%%%%%%%%%%%%%%%%%%%%%%%%%%%%%%%%%%%%
%%%%%%%%%%%%%%%%%%%%%%%%%%%%%%%%%%%%%%%%%%%%%%%%%%%%%%%%%%%%
\begin{abstract} 
Spacetime curvature plays the primary role in general relativity 
but Einstein later considered a theory where torsion was the central 
quantity.  Just as the Einstein-Hilbert action in the Ricci curvature 
scalar $R$ can be generalized to $f(R)$ gravity, we consider extensions 
of teleparallel, or torsion scalar $T$, gravity to $f(T)$ theories.  
The field equations are naturally second order, avoiding pathologies, 
and can give rise to cosmic acceleration with unique features. 
\end{abstract} 

\maketitle

%%%%%%%%%%%%%%%%%%%%%%%%%%%%%%%%%%%%%%%%%%%%%%%%%%%%%%%%%%%%
%%%%%%%%%%%%%%%%%%%%%%%%%%%%%%%%%%%%%%%%%%%%%%%%%%%%%%%%%%%%
\section{Introduction \label{sec:intro}}

Acceleration of the cosmic expansion is one of the premier 
mysteries of physics and may be the clearest clue to properties 
of gravity beyond general relativity.  Extensions to gravity 
have been considered by making the action a function of the 
spacetime curvature scalar $R$ or other curvature invariants, 
by coupling this Ricci scalar to a scalar field, by introducing 
a vector field contribution, and by using properties of gravity 
in higher dimensional spacetimes.  Here we take a wholly 
different path, avoiding the curvature completely, although our 
results will end up with interesting relations to each of the above 
mentioned theories. 

Rather use the curvature defined via the Levi-Civita connection, 
one could explore the opposite approach and use the Weitzenb{\"o}ck 
connection that has no curvature but instead torsion.  This has the 
interesting property that the torsion is formed wholly from products 
of first derivatives of the tetrad, with no second derivatives appearing 
in the torsion tensor.  In fact, this approach was taken by Einstein in 
1928 \cite{ein28,eineng}, under the name ``Fern-Parallelismus'' or ``distant 
parallelism'' or ``teleparallelism''.  It is closely related to standard 
general relativity, differing only in terms involving total derivatives 
in the action, i.e.\ boundary terms. 

In this paper, we investigate extensions where a scalar formed from 
contractions of the torsion tensor is promoted to a function of that 
scalar.  This parallels the concept of extension of the Ricci scalar $R$ 
in the Einstein-Hilbert action to a function $f(R)$, which has attracted 
much attention in recent years as a way to explain acceleration of the 
universe \cite{durrer,sotiriou}.  The generalized $f(T)$ torsion theory 
has the advantage, however, of keeping its field equations second order 
due to the lack of second derivatives, unlike the fourth order equations 
(at least in the metric formulation) of $f(R)$ theory that can lead to 
pathologies.

%%%%%%%%%%%%%%%%%%%%%%%%%%%%%%%%%%%%%%%%%%%%%%%%%%%%%%%%%%%%%%%%%%%%%
\section{Cosmological Equations} \label{sec:cos} 

We start with the Robertson-Walker metric for a homogeneous and isotropic 
space with zero spatial curvature: 
\beq 
ds^2=-dt^2+a^2(t)\,\delta_{ij} dx^i dx^j\,, 
\eeq 
where $a$ is the expansion factor. 
The orthonormal tetrad components $e_A(x^\mu)$ relate to the metric through 
\beq 
g_{\mu\nu}=\eta_{AB}e^A_\mu e^B_\nu\,, 
\eeq 
where $A$, $B$ are indices running over 0, 1, 2, 3 for the tangent space of 
the manifold and $\mu$, $\nu$ are coordinate indices on the manifold, also 
running over 0, 1, 2, 3 (with $i$, $j$ being the spatial indices). 

The torsion tensor and permutations (note the different symmetry properties 
from the curvature case), are 
\beqa 
T^\rho{}_{\mu\nu}&\equiv&-e_A^\rho\,(\partial_\mu e^A_\nu-\partial_\nu 
e^A_\mu)\\ 
K^{\mu\nu}{}_\rho&\equiv&-\frac{1}{2}(T^{\mu\nu}{}_\rho-T^{\nu\mu}{}_\rho  
-T_\rho{}^{\mu\nu})\\ 
S_\rho{}^{\mu\nu}&\equiv&\frac{1}{2}(K^{\mu\nu}{}_\rho+\delta^\mu_\rho 
T^{\alpha\nu}{}_\alpha-\delta^\nu_\rho T^{\alpha\mu}{}_\alpha)\,. 
\eeqa 
In place of the Ricci scalar for the Lagrangian density, one has the 
torsion scalar (also see \cite{beng}) 
\beq 
T\equiv S_\rho{}^{\mu\nu} T^\rho{}_{\mu\nu}\,, \label{eq:tdef} 
\eeq 
and the gravitational action is 
\beq 
I=\frac{1}{16\pi G}\int d^4 x\,|e|\,T\,, 
\eeq 
where $|e|=\det(e^A_\mu)=\sqrt{-g}$. 
For a more detailed derivation giving a clearer picture of the relation 
to general relativity, with the difference arising in boundary terms, 
see \cite{hehl,hayash1,hayash2,eanna}. 

Following \cite{beng} we now promote $T$ to a function, replacing it 
in the action by $T+f(T)$, in analogy to $f(R)$ gravity (see, e.g., 
\cite{durrer,sotiriou}).  The modified Friedmann equations of motion 
are (cf.~\cite{beng} with different notation) 
\beqa 
H^2&=&\frac{8\pi G}{3}\rho -\frac{f}{6}-2H^2 f_T\label{eq:fried1}\\ 
(H^2)'&=&\frac{16\pi GP+6H^2+f+12H^2 f_T}{24H^2 f_{TT}-2-2f_T}
\,, \label{eq:fried2} 
%[24H^2 f_{TT}-2f_T](H^2)'-12H^2 f_T-f&=&16\pi GP\,, \label{eq:fried2}
\eeqa 
where the Hubble expansion parameter $H=\dot a/a$, a prime denotes 
a derivative with respect to $\ln a$, $\rho$ is the energy density and 
$P$ is the pressure.  Evaluating Eq.~(\ref{eq:tdef}) for the unperturbed 
metric, one finds $T=-6H^2$, so one can use $T$ and $H$ interchangeably. 

Taking a universe with only matter (so $P=0$), we find the 
solution $T(a)$ in closed form: 
\beq 
a(T)=\exp\left\{-\frac{1}{3}\int_{-6H_0^2}^T \frac{d\tilde T}{\tilde T}\, 
\frac{1+f_T+2{\tilde T}f_{TT}}{1-f/\tilde T+2f_T}\right\}\,. 
\eeq 
One can also define an effective dark energy density and equation of state 
\beqa 
\rho_{de}&=&\frac{1}{16\pi G}\,(-f+2Tf_T)\\ 
w&=&-1+\frac{1}{3}\frac{T'}{T}\,\frac{f_T+2Tf_{TT}}{f/T-2f_T} 
=-\frac{f/T-f_T+2Tf_{TT}}{(1+f_T+2Tf_{TT})(f/T-2f_T)}\,. 
%=-1-\frac{1-f/T+2f_T}{1+f_T+2Tf_{TT}}\,\frac{f_T+2Tf_{TT}}{f/T-2f_T}\,. 
\eeqa 

From the modified Friedmann equations (\ref{eq:fried1}), (\ref{eq:fried2}) 
we see that a constant $f$ acts just like a cosmological constant, and 
$f$ linear in $T$ (i.e.\ $f_T=$constant) is simply a redefinition of 
Newton's constant $G$.

%%%%%%%%%%%%%%%%%%%%%%%%%%%%%%%%%%%%%%%%%%%%%%%%%%%%%%%%%%%%%%%%%%%%%
\section{Results for Cosmic Acceleration} \label{sec:forms} 

At high redshift, we desire general relativity to hold so as to 
agree with primordial nucleosynthesis and cosmic microwave background 
constraints.  Therefore we want $f/T\to0$ at early times, $a\ll1$. 
Regarding the future, Eq.~(\ref{eq:fried2}) says an asymptotic future 
de Sitter state (with $H=$ constant and $w=-1$), for example, occurs when 
the numerator (but not the denominator) vanishes.  Many functions 
$f(T)$ can give a de Sitter fate for the universe; here we examine two 
models. 

As a first example, consider a power law 
\beq 
f=\alpha (-T)^n=\alpha\,6^n H^{2n}\,. \label{eq:Tn}
\eeq 
From Eq.~(\ref{eq:fried1}), the dimensionless matter density today 
\beq 
\om=\frac{8\pi G\rho_m(a=1)}{3H_0^2}=1+\frac{f(T_0)}{6H_0^2}+2f_T\,, 
\eeq 
so $\alpha=(6H_0^2)^{1-n}(1-\om)/(2n-1)$.  The Hubble expansion freezes 
in the future at 
the value $H_\infty=H_0\,(1-\om)^{1/[2(1-n)]}$.  The effective dark energy 
equation of state varies from $w=-1+n$ in the past to $w=-1$ in the future. 
For example, solving the modified Friedmann equations numerically, for 
$n=0.25$ one has $w_0=-0.91$ and $w(a=0.5)=-0.81$; for this form of 
$f(T)$ to be a viable model compared to current data one needs $n\ll1$. 

Such a functional form as Eq.~(\ref{eq:Tn}) results in a power of $H$ 
being added to the Friedmann equation and is equivalent (at least at the 
background level) to the phenomenological models of 
\cite{dvaliturner,cardassian}.  In \cite{paths} it was shown that such 
models behave as freezing scalar fields, and in particular approach a 
de Sitter state in the future along the curve $w'=3w(1+w)$.  We have 
numerically solved the equations of motion to verify that this holds 
for such an $f(T)$ as well.  Note that $n=1/2$ gives the same 
expansion history as DGP gravity \cite{dgp,ddg}, so $f(T)$ gravity can 
be viewed as having some connection to higher dimension theories. 

Another fine tuning for the power law models is that one has a similar 
condition to $f(R)$ gravity in that the factor $f_T$, acting to rescale 
Newton's constant, should be small.  The condition is not as sensitive 
as in $f(R)$ gravity, because there $R$ changes with scale so solar 
system and galactic constraints impose tight bounds on $f_R$.  For 
$f(T)$ theories, $T$ is much less scale dependent (of order $(k/H)^2\Phi^2$  
\cite{smithlin}) so the time variation of $G$ gives the main limit. 
This again imposes $n\ll1$. 

To keep the variation of the gravitational coupling small within $f(R)$ 
theory, \cite{linfr} adopted an exponential dependence on the curvature 
scalar.  Here we explore a similar exponential dependence on the torsion 
scalar as an example.  We take the form 
\beq 
f=c T_0\,\left(1-e^{-p\sqrt{T/T_0}}\right)\,, \label{eq:ftexp}
\eeq 
where $c=(1-\om)/[1-(1+p)e^{-p}]$.  
Note that there is only one parameter, $p$, besides the value of the 
matter density today, $\om$, and the functional form is exponential 
in $H/H_0$.  

\begin{figure}[!htb]
\begin{center}
\psfig{file=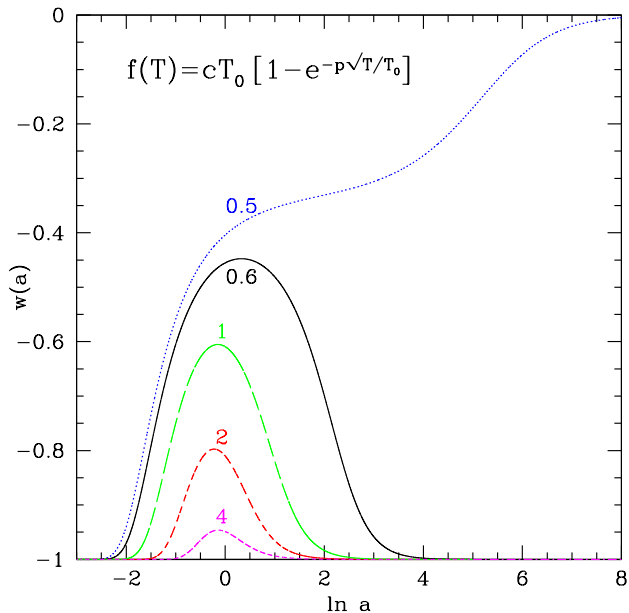,width=3.4in}
\caption{Effective dark energy equation of state is plotted vs.~scale 
factor for the exponential $f(T)$ model of Eq.~(\ref{eq:ftexp}).  Curves 
are labeled with the value of the one free parameter $p$.  The 
model acts like a cosmological constant at high redshift and in the future, 
except for $p<0.51$ it behaves as Einstein-de Sitter in the future. 
}
\label{fig:fexpwa}
\end{center}
\end{figure}

Figure~\ref{fig:fexpwa} illustrates the behavior of the equation of 
state for several values of $p$.  At high redshift the model acts like 
$\Lambda$CDM, then it deviates to $w>-1$ and is asymptotically 
attracted to a de Sitter fate (for $p>0.51$, otherwise the asymptotic 
equation of state is $w=0$).  The parameter $p$ mainly controls the 
amplitude of the deviation from $w=-1$.  No fine tuning is needed, with 
$p\gtrsim3$ allowed by current cosmological observations of the expansion 
history.  (Note that the different model used in version 1 of this paper 
had a hidden cosmological constant and so is not of interest; it also 
does not cross $w=-1$, as pointed out by \cite{wuyu}.  Models acting 
like $\Lambda$CDM at high redshift have difficulty crossing $w=-1$.)

%%%%%%%%%%%%%%%%%%%%%%%%%%%%%%%%%%%%%%%%%%%%%%%%%%%%%%%%%%%%%%%%%%%%%
\section{Summary and Conclusions} \label{sec:concl} 

The class of $f(T)$ gravity theories is an intriguing generalization 
of Einstein's ``new general relativity'', taking a curvature-free 
approach and instead using a connection with torsion.  It is analogous 
to the $f(R)$ extension of the Einstein-Hilbert action of standard 
general relativity, but has the advantage of second order field 
equations.  We have also seen that it can be related to the form of 
modifications to the Friedmann equations due to higher dimensional gravity 
theories such as DGP. 

It is also related to scalar-tensor gravity.  Writing the gravitational 
action as 
\beq 
S=\int d^4x\,|e|\,\{T+f(T)+(T-A)\,[1+f_A(A)]\}\,, 
\eeq 
one can view the last term as a Lagrange multiplier term and find 
an equivalent scalar-tensor theory with $A=T$ and an effective potential 
\beqa 
V_{\rm eff}(\psi)&=&\frac{T}{1+f_T}-\frac{T+f}{(1+f_T)^2} \\ 
\psi&=&-\ln(1+f_T)\,. 
\eeqa 
Furthermore, Einstein originally introduced teleparallelism to obtain a 
vector field component of the field equations \cite{ein25,eineng}, 
intending to unify gravity and electromagnetism.  Recently, interest has 
grown in vector fields, ``Einstein aether theories'', as a way to obtain 
cosmic acceleration \cite{vectorde}.  These theories can also give 
modifications to the field equations involving functions of $H^2$, i.e.\ 
$T$ (see, e.g., \cite{zuntz}).  Indeed they can be viewed as closely 
related to torsion theories (see \cite{hayash1,hayash2} for details). 

Thus, torsion theories can unify a number of interesting extensions of 
gravity beyond general relativity.  In investigating the nature of 
gravitation, we may find that Einstein presaged the acceleration of the 
universe not only through the cosmological constant but through a 
generalization of ``Einstein's other gravity''.

%%%%%%%%%%%%%%%%%%%%%%%%%%%%%%%%%%%%%%%%%%%%%%%%%%%%%%%%%%%%%%%
\acknowledgments 

I thank the Aspen Center for Physics for hospitality and a valuable 
program on gravity, and Tristan Smith for gravilicious discussions. 
This work has been supported in part by the Director, Office of Science, 
Office of High Energy Physics, of the U.S.\ Department of Energy under 
Contract No.\ DE-AC02-05CH11231, and the World Class University grant 
R32-2009-000-10130-0 through the National Research Foundation, Ministry 
of Education, Science and Technology of Korea.

\end{document}